\documentclass{article}

\usepackage[preprint, nonatbib]{neurips_2020}

\usepackage[utf8]{inputenc} 
\usepackage[T1]{fontenc}    
\usepackage{hyperref}       
\usepackage{url}            
\usepackage{booktabs}       
\usepackage{amsfonts}       
\usepackage{nicefrac}       
\usepackage{microtype}      
\usepackage{caption}
\usepackage{subcaption}
\usepackage{graphicx}
\usepackage[load-configurations=version-1]{siunitx} 

\newcommand{\boldentry}[2]{%
  \multicolumn{1}{S[table-format=#1,
                    mode=text,
                    text-rm=\fontseries{b}\selectfont
                   ]}{#2}}

\title{Self-Supervision Closes the Gap Between Weak and Strong Supervision in Histology}

\author{%
  Olivier Dehaene\\
  Owkin Inc.\\
  \texttt{olivier.dehaene@owkin.com} \\
   \And
   Axel Camara \\
   Owkin Inc.\\
   \texttt{axel.camara@owkin.com} \\
   \AND
   Olivier Moindrot \\
   Owkin Inc.\\
   \texttt{olivier.moindrot@owkin.com} \\
   \And
   Axel de Lavergne \\
   Owkin Inc.\\
   \texttt{axel.de-lavergne@owkin.com} \\
   \And
   Pierre Courtiol \\
   Owkin Inc.\\
   \texttt{pierre.courtiol@owkin.com} \\
}

\begin{document}

\maketitle

\begin{abstract}
One of the biggest challenges for applying machine learning to histopathology is weak supervision: whole-slide images have billions of pixels yet often only one global label. The state of the art therefore relies on strongly-supervised model training using additional local annotations from domain experts. However, in the absence of detailed annotations, most weakly-supervised approaches depend on a frozen feature extractor pre-trained on ImageNet. We identify this as a key weakness and propose to train an in-domain feature extractor on histology images using MoCo v2, a recent self-supervised learning algorithm. Experimental results on Camelyon16 and TCGA show that the proposed extractor greatly outperforms its ImageNet counterpart. In particular, our results improve the weakly-supervised state of the art on Camelyon16 from 91.4\% to 98.7\% AUC, thereby closing the gap with strongly-supervised models that reach 99.3\% AUC. Through these experiments, we demonstrate that feature extractors trained via self-supervised learning can act as drop-in replacements to significantly improve existing machine learning techniques in histology. Lastly, we show that the learned embedding space exhibits biologically meaningful separation of tissue structures.
\end{abstract}

\section{Introduction}
\label{sec:introduction}

Histopathology is the gold standard for diagnosis in many diseases, and especially in oncology.
One example of a routine task performed by pathologists is metastasis detection in hematoxylin and eosin (H\&E) stained slides of lymph nodes, which can prove challenging as whole-slide images  (WSI) can be over 25 mm wide while micro-metastases can be as small as 50 ${\mu}m$ wide \cite{bird2009detection}.
However, in most existing cohorts, WSIs are only associated with a pathology report containing global labels, but that lacks detailed pixel-level annotations.
In the example above a lymph node slide would only have a binary slide-level label indicating whether it contains any metastasis.

Despite these limitations, recent work has proved that machine learning can perform well on histology images \cite{Bulten_2020}, sometimes as well as human experts \cite{Campanella2019Clinical-gradeImages}.
Models that attempt to learn directly from global labels belong to the field of \textit{weakly-supervised learning}, and have to propagate the slide-level information down to individual pixels of the WSI. 
Other models rely instead on additional pixel-level annotations and are referred as \textit{strongly-supervised}, however these pathologist annotations are costly and time-consuming.

By taking advantage of the added information from annotations, strongly-supervised models perform better than weakly-supervised ones. On Camelyon16 \cite{Bejnordi2017DiagnosticCancer}, the state of the art for breast cancer metastasis detection is strongly-supervised and reaches an area under the ROC curve (AUC) of 99.3\% \cite{WangDeepCancer}.
Without annotations and despite many recent advances, weakly-supervised learning on Camelyon16 only achieves 91.4\% AUC \cite{Lu2020Semi-supervisedPresentation}. 
However, other research questions in histology consist in predicting a label that is not directly visible in the slide, such as the overall survival of the patient \cite{yao2020whole} or gene expression from transcriptomic data \cite{he2rna}.
In these situations, annotations can provide helpful regions of interest but are not enough to solve the task itself, so weakly-supervised learning remains the only option and is therefore a key method for machine learning in histology.

Another challenge for weak supervision is that current constraints on GPU memory prevent full end-to-end learning, as the input WSI itself already takes up to 8GB \cite{Pinckaers2020StreamingImages}.
Most approaches thus rely on a pre-processing step to reduce the dimensionality of the input, often separating the slide into smaller images (tiles) on which conventional architectures, such as convolutional neural network, can operate.
This network, the feature extractor, can be either fine-tuned \cite{Coudray2018ClassificationLearning, Kather2019DeepCancer} or frozen \cite{Ilse2018Attention-basedLearning, Courtiol2018ClassificationApproach}, but in both cases, the network is often pre-trained on ImageNet \cite{DengImageNet:Database}, an \textit{out-of-domain} dataset of natural images.

We identify the use of out-of-domain pre-training as the main weakness in current weakly-supervised methods.
In this paper, we propose to train a new \textit{in-domain} encoder on histology tiles using unsupervised learning.
Recent advances \cite{Chen2020ARepresentations, GrillBootstrapLearning} have significantly improved the transfer learning performance of unsupervised pre-training.
We use one such state-of-the-art self-supervised algorithm, MoCo v2 \cite{ChenImprovedLearning}, to train a tile-level feature extractor without supervision.
This approach is very well suited to histology, since each WSI contains tens of thousands of unlabeled tiles.

Our work makes the following contributions:
\begin{itemize}
\item  In Section \ref{sec:methods}, we propose a generic pipeline combining self-supervised pre-training and weakly-supervised learning.
\item We show in Sections \ref{sec:experiments_camelyon} and \ref{sec:experiments_cms} that our approach is generic enough to yield significant improvements on two datasets for three weakly-supervised algorithms.
In particular we close most of the gap between weak and strong supervision on Camelyon16 by reaching 98.7\% AUC on the competition test set.
In Section \ref{sec:experiments_interpretability}, we bring out the biological relevance of the self-supervised learned features.
\item Finally, we demonstrate in Section \ref{sec:experiments_transfer_learning} that self-supervised feature extractors can transfer across different organs and tumor types.
\end{itemize}

\section{Related Work}
\label{sec:related_work}

\begin{figure}[htbp]
\centering

     \begin{subfigure}[b]{0.45\textwidth}
         \centering
         \includegraphics[width=\textwidth]{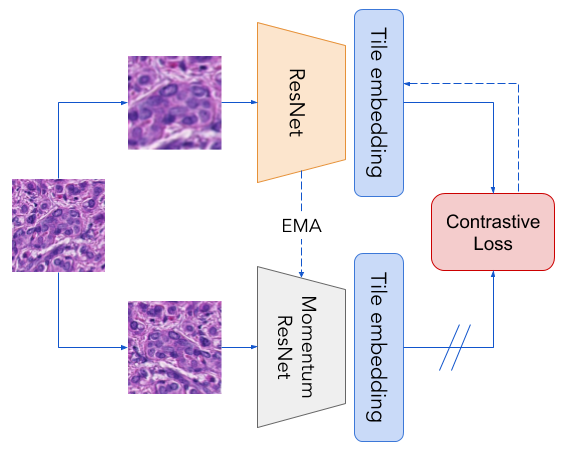}
         \caption{}
         \label{fig:archi_1}%
     \end{subfigure}
     \hfill
     \begin{subfigure}[b]{0.45\textwidth}
         \centering
         \includegraphics[width=\textwidth]{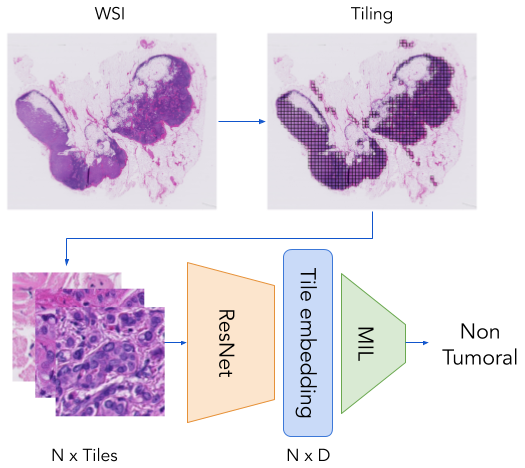}
         \caption{}
         \label{fig:archi_2}%
     \end{subfigure}
  
\caption{Proposed pipeline. We train a ResNet encoder on histology tiles using MoCo v2, a self-supervised learning algorithm (a) and use the trained encoder as a feature extractor for multiple instance learning (MIL) (b). More details in Section \ref{sec:methods}.}
\label{fig:architecture}
\end{figure}

\paragraph{Strong supervision} A number of previous works in histology have relied on pathologists hand-crafted pixel-level annotations to train machine learning models \cite{cirecsan2013mitosis,chan2019histosegnet,saltz2018spatial}.
For tasks like cancer detection, the slide-level label is derived from local tumor annotations: if no tumor is found, the slide-level label is negative.
For other tasks, such as molecular subtype prediction, the global label is obtained from external data, such as genomic sequencing, and cannot be expressed as a purely local information.
However, in Kather et al. \cite{Kather2019DeepCancer} and Sirinukunwattana et al. \cite{Sirinukunwattana2020Image-basedLearning}, expert knowledge is added to the model by annotating regions of interest (the tumor area), which may be seen as a weaker form of strong supervision.

\paragraph{Weak supervision} When annotations are not available, models are trained with weakly-supervised learning using only the slide-level labels as supervision. One of the most common approaches frames the problem as multiple instance learning (MIL) \cite{Ilse2018Attention-basedLearningb}.
Weakly-supervised models have been used to predict patient survival in hepatocellular carcinoma \cite{saillard2020predicting} or microsatellite instability in colorectal cancer \cite{echle2020clinical}.

\paragraph{End-to-end training on WSI}
GPU memory limitations prevent full end-to-end learning on WSIs, and most weakly-supervised approaches use a frozen feature extractor to reduce the input dimension.
Pinckaers et al. \cite{Pinckaers2020DetectionLabels,Pinckaers2020StreamingImages} used a streaming neural network to permit end-to-end training, but at the cost of only operating on a down-sampled slide 16,384 pixels wide.
Campanella et al. \cite{Campanella2019Clinical-gradeImages} used the MIL framework and managed to train their network end-to-end by back-propagating only on the top tiles.
Coudray et al. \cite{Coudray2018ClassificationLearning} directly used slide-level labels as local tile labels to train their network.
Other papers have tried several techniques to pre-train the feature extractor including supervised multitask learning \cite{Tellez2019NeuralAnalysis} or  Bidirectional Generative Adversarial Networks \cite{DBLP:journals/corr/abs-1811-02840}.

\paragraph{Self-supervised algorithms} Over the past two years, rapid developments have been made in the field of self-supervised learning \cite{ChenBigLearners, caron2020unsupervised}.
He et al. \cite{He2019MomentumLearning} showed that self-supervised pre-training outperformed supervised pre-training on ImageNet for downstream transfer learning performance.
Among the latest self-supervised algorithms, MoCo v2 \cite{ChenImprovedLearning} can work with smaller batch sizes which makes it a better fit for lower computational resources.
Lu et al. \cite{Lu2020Semi-supervisedPresentation,Lu2019Semi-SupervisedCoding} applied Contrastive Predictive Coding \cite{DBLP:journals/corr/abs-1807-03748} to histology to pre-train their feature extractor.

\section{Methods}
\label{sec:methods}

Our proposed pipeline is summarized in Figure \ref{fig:architecture}.
In Section \ref{sec:methods_histo_workflow}, we detail the standard workflow for weakly-supervised learning in histology.
We then describe in Section \ref{sec:methods_ssl} a generic method to pre-train the feature extractor with self-supervised learning, which can be applied to improve performance in most existing models in histology.

\subsection{Weakly-supervised learning in histopathology}
\label{sec:methods_histo_workflow}

To train deep learning models on histology images with only slide-level labels, we use a typical three-step workflow illustrated in Figure \ref{fig:archi_2}: tiling, feature extraction and multiple instance learning training.

\paragraph{Tiling} 
We extract from tissue regions a grid of $N$ tiles of fixed size at a set zoom level, e.g. $0.5$ microns per pixel.
The total number of tiles $N$ depends on the tissue area in the WSI.

\paragraph{Feature extraction} 
A frozen feature extractor, such as ResNet50 \cite{he2015deep} pre-trained on ImageNet, is applied to each of the $N$ tiles as a pre-processing step.
This encoder extracts $D$ relevant features from each tile.
We therefore obtain a matrix of size $N \times D$ for each slide.

\paragraph{Multiple Instance Learning}
The training phase begins at this step, with an $N \times D$ matrix as input and a slide-level label as output.
In our paper we consider three recent approaches following the multiple instance learning (MIL) framework: Weldon \cite{DurandWELDON:Networks}, Chowder \cite{Courtiol2018ClassificationApproach} and DeepMIL \cite{Ilse2018Attention-basedLearning}.
One common attribute of these architectures is that they all first compute an attention score for each tile of the WSI.
In Weldon and Chowder, only the minimum and maximum scores are kept and combined into a final prediction.
DeepMIL uses a weighted sum of the tile features using the tile attention scores and computes the prediction using that average representation.
Implementation details can be found in Appendix \ref{apd:mil}.

\subsection{Training a better feature extractor with self-supervised learning}
\label{sec:methods_ssl}

\renewcommand{\thefootnote}{\fnsymbol{footnote}}

\begin{table}[htbp!]
  \centering
  \caption{Performance for tumor detection (AUC percentage) on Camelyon16 in 5-fold cross-validation (repeated 5 times) on the train set and on the competition test set (5 independent runs). Using the MoCo v2 feature extractor for weakly-supervised learning closes most of the gap with the strongly-supervised state of the art.}
  \label{tab:cam16}
  {\begin{tabular}{lcS[table-format=1.1(2)]S[table-format=1.1(2)]}
  \toprule 
  \bfseries Method & \bfseries \shortstack{Feature Extractor\\Training Method} & \bfseries \shortstack{Train Cross-\\Validation} & \bfseries \shortstack{Competition\\ Test Set} \\
  \midrule
  \textit{Strongly-Supervised} \cite{WangDeepCancer}  & - & {-} & \boldentry{1.1(2)}{99.3} \\
  \midrule
  CLAM \cite{Lu2020Semi-supervisedPresentation}  & CPC & {-}  & 91.4(23) \\
  \midrule
  Weldon & ImageNet & 76.9(166) & 65.3(137)  \\
  Chowder &   & 82.3(154) & 79.6(51) \\
  DeepMIL &  & 88.7(47) & 82.9(20)  \\
  \midrule
  Weldon & MoCo V2 & 97.9(14) & 98.3(7)  \\
  Chowder &  & \boldentry{1.1(2)}{98.3(12)}  & 97.7(5) \\
  DeepMIL &  & 96.3(21)  & \boldentry{1.1(2)}{98.7(2)} \\
  \bottomrule
  \end{tabular}}
\end{table}

As stated before, one key limitation of the approach described in Section \ref{sec:methods_histo_workflow} is the use of a feature extractor trained on out-of-domain data.
Here we propose applying recent advances in self-supervised learning to train a better in-domain feature extractor on histology tiles, without annotations.

We use Momentum Contrast v2 (MoCo v2) \cite{ChenImprovedLearning}, a self-supervised learning algorithm using contrastive loss to shape an embedding space where different augmented views of the same image are close together.
MoCo v2 adds incremental improvements on the original MoCo paper \cite{He2019MomentumLearning}.
As shown in Figure \ref{fig:archi_1}, a batch of tile goes through two ResNet encoders with different data augmentations.
The contrastive loss then pushes pairs of matching tiles closer, and pairs of different tiles apart.
Gradients are only back-propagated through the first of these two encoders.
The second encoder is updated with an exponential moving average (EMA) of the first encoder's weights.

To apply this self-supervised framework on histology data, we concatenate tiles from all WSIs extracted at the tiling step to form a training dataset.
We then train the feature extractor with MoCo v2 on this set of unlabeled tile images.
We only modify the data augmentation scheme by adding $\ang{90}$ rotations and vertical flips, since histology tiles contain the same information regardless of their orientation.
Implementation details can be found in Appendix \ref{apd:moco}.

\section{Experiments}
\label{sec:experiments}

In this section, we evaluate the performance of our approach on two datasets (Sections \ref{sec:experiments_camelyon} and \ref{sec:experiments_cms}), look at the interpretability of the learned representation (Section \ref{sec:experiments_interpretability}) and its transfer learning potential (Section \ref{sec:experiments_transfer_learning}).

\subsection{Metastasis detection on Camelyon16}
\label{sec:experiments_camelyon}

The Camelyon16 challenge consists in automatically detecting breast cancer metastases in sentinel lymph node WSIs.
AUC for 5-fold cross-validation (repeated 5 times) on the train set and AUC on the hold-out test set (5 independent runs) are presented in Table \ref{tab:cam16}.

\paragraph{Dataset} The Camelyon16 dataset contains a total of 400 sentinel lymph node H\&E-stained WSIs from two medical centers, split into 270 for training and 130 for testing.
Local annotations done by pathologists at pixel-level for the slides containing metastases are provided.

\paragraph{Baselines} In Wang et al. \cite{WangDeepCancer}, 
a patch-based classifier is trained using local tile annotations and tile predictions are combined in a post-processing stage to establish a strongly-supervised state of the art at 99.3\% AUC on the test set.
On the other hand, the weakly-supervised learning state of the art achieves 91.4\% AUC by pre-training the encoder with Contrastive Predictive Coding \cite{Lu2020Semi-supervisedPresentation}.

\paragraph{Results}
In our experiments we show that switching from a feature extractor trained on ImageNet to one trained with MoCo v2 on Camelyon16 tiles improves results significantly.
We use the exact same parameters for both experiments, only replacing the feature extractor itself. 
Our best model obtains 98.7\% AUC on the test set, setting a new state of the art for weakly-supervised learning on Camelyon16 (+7.3 AUC points compared to Lu et al. \cite{Lu2020Semi-supervisedPresentation}), which closes the gap with the strongly-supervised baseline at 99.3\% AUC.

We empirically observe performance instability across folds when training Weldon and Chowder on ImageNet features, with standard deviations of 16.6\% and 15.4\% AUC respectively (Table \ref{tab:cam16}).
Courtiol et al. \cite{Courtiol2018ClassificationApproach} reduced the variability by using an ensemble of 50 models.
In our experiments, the standard deviation is divided by a factor of 10 when switching to MoCo v2 features, which demonstrates the robustness and quality of this representation.

\subsection{Consensus Molecular Subtype classification in colorectal cancer}
\label{sec:experiments_cms}

\renewcommand{\thefootnote}{\fnsymbol{footnote}}

\begin{table}[htbp]
\centering

\caption{Performance for Consensus Molecular Subtype (CMS) classification (one vs. all AUC percentage for each subtype, and macro average on all subtypes) on TCGA-COAD in 5-fold cross-validation (repeated 5 times). Changing the ImageNet feature extractor to MoCo v2 improves results substantially. Our pipeline obtains comparable results to the state of the art without using tumor annotations, ensembling and additional slides.}
\label{tab:coad}
    \addtolength{\leftskip} {-2cm}
    \addtolength{\rightskip}{-2cm}
    \begin{tabular}{lcS[table-format=1.1(3)]S[table-format=1.1(3)]S[table-format=1.1(3)]S[table-format=1.1(3)]S[table-format=1.1(3)]}
  \toprule
  \bfseries Method & \bfseries \shortstack{Feature Extractor\\ Training Method} & \bfseries CMS1 & \bfseries CMS2 & \bfseries CMS3 & \bfseries CMS4 & \bfseries \shortstack{Macro\\ Average} \\
  \midrule
  imCMS \cite{Sirinukunwattana2020Image-basedLearning} \footnotemark[1] & Tile Classification  & 85(5) & \boldentry{1.1(3)}{89(3)} & 78(7) & \boldentry{1.1(3)}{83(4)} &  \boldentry{1.1(3)}{84(3)} \\
   & with Tumor Annotations \\
  DeepHistology \cite{Kather2020Pan-cancerAlterations} & ImageNet & 70 & 69 & 66 & 60 & 66.2 \\
      \midrule
  Weldon & ImageNet & 74.8(72) & 68.9(51) &  66.9(84) &  63.9(71) &  68.6(43)  \\
  Chowder &  & 75.1(70) & 70.9(58) &  64.2(76) &  63.4(65) &  68.4(48)  \\
  DeepMIL &  & 77.4(61) & 74.6(57) &  73.9(63) &  62.5(74) &  72.1(40)  \\
  \midrule
  Weldon & MoCo V2 & 87.7(47) & 81.9(50) &  \boldentry{1.1(3)}{82.9(50)} &  68.2(59) &  80.1(33)  \\
  Chowder &  & 86.9(54) & 79.9(50) &  79.5(64) &  66.9(66) &  78.3(36)  \\
  DeepMIL &  & \boldentry{1.1(3)}{88.2(44)} & 82.7(40) &  80.3(58) &  68.8(62) &  80.0(35)  \\
  \bottomrule
  \end{tabular}
\end{table}

\footnotetext[1]{Ensemble of 5 models.}

In this section, we describe results on Consensus Molecular Subtype (CMS) classification on colorectal cancer (Table \ref{tab:coad}). The AUC is computed with 5-fold cross-validation, repeated 5 times.
All models are trained with multiclass classification, and we report one vs. all AUC for each category, as well as the average AUC (Macro Average).

\paragraph{Dataset} The TCGA-COAD dataset contains a total of 461 colorectal cancer WSIs. Among these, 364 are classified in one of the four transcriptome-based consensus molecular subtypes (CMS) \cite{Guinney2015TheCancer}.

\paragraph{Baselines}
Sirinukunwattana et al. \cite{Sirinukunwattana2020Image-basedLearning} used additional tumor annotations to train a model called \textit{imCMS} only on tumoral tiles.
Although these annotations are not directly linked to the CMS labels, they help the model to focus on meaningful regions of the WSI. 
Their model is trained on FOCUS, an external cohort of 510 slides and transfered to TCGA-COAD with Domain Adaptation techniques.
Kather et al. \cite{Kather2020Pan-cancerAlterations} used a weakly-supervised approach called \textit{DeepHistology} that doesn't use any annotation, but obtain lower results.

\paragraph{Results}
Similarly to results on Camelyon16, switching from ImageNet pre-training to an encoder trained on TCGA-COAD tiles with MoCo v2 increases the AUC by an average of 10 AUC points.
Both Weldon and DeepMIL models outperform the state-of-the-art results of \textit{imCMS} in CMS1 (88.2\% AUC) and CMS3 (82.9\% AUC) subtypes, while using less data, no tumor annotations and no ensembling.

On CMS4, our approach succeeds in increasing the AUC but remains far from the results of Sirinukunwattana et al. \cite{Sirinukunwattana2020Image-basedLearning}. The high infiltration of stromal cells within CMS4 tumors \cite{Guinney2015TheCancer} makes it more difficult for the model to focus on the tumor regions.
Without local tumor annotations, our weakly-supervised method therefore struggles to accurately detect CMS4 cases.

\subsection{Interpretability}
\label{sec:experiments_interpretability}

\begin{figure}[htbp]
\makebox[\textwidth][c]{
\centering
     \begin{subfigure}[b]{0.35\textwidth}
         \centering
         \includegraphics[height=\textwidth]{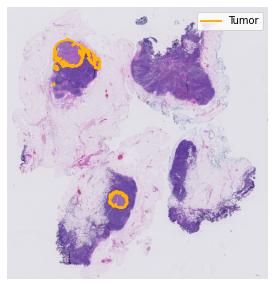}
         \caption{}
     \end{subfigure}
     \hfill
     \begin{subfigure}[b]{0.35\textwidth}
         \centering
         \includegraphics[height=\textwidth]{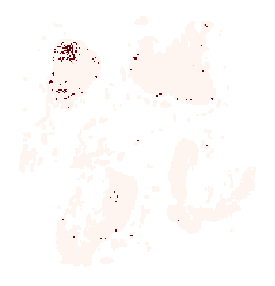}
         \caption{}
     \end{subfigure}
     \hfill
     \begin{subfigure}[b]{0.35\textwidth}
         \centering
         \includegraphics[height=\textwidth]{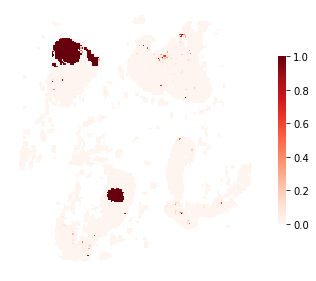}
         \caption{}
         \label{fig:cam_test_001_moco_heatmap}
     \end{subfigure}
}
  \caption{(a) Tumor annotations (orange) displayed on a Camelyon16 test slide. (b) The best performing cluster on ImageNet features among 10 clusters obtains 69.4\% AUC. (c) The best performing cluster on MoCoV2 features among 10 clusters obtains 95.1\% AUC and matches almost perfectly the annotations, while being fully unsupervised.}
\label{fig:camelyon_cluster}
\end{figure}

\begin{figure}[htbp]
\makebox[\textwidth][c]{

\centering

     \begin{subfigure}[b]{0.75\textwidth}
         \centering
         \includegraphics[width=\textwidth]{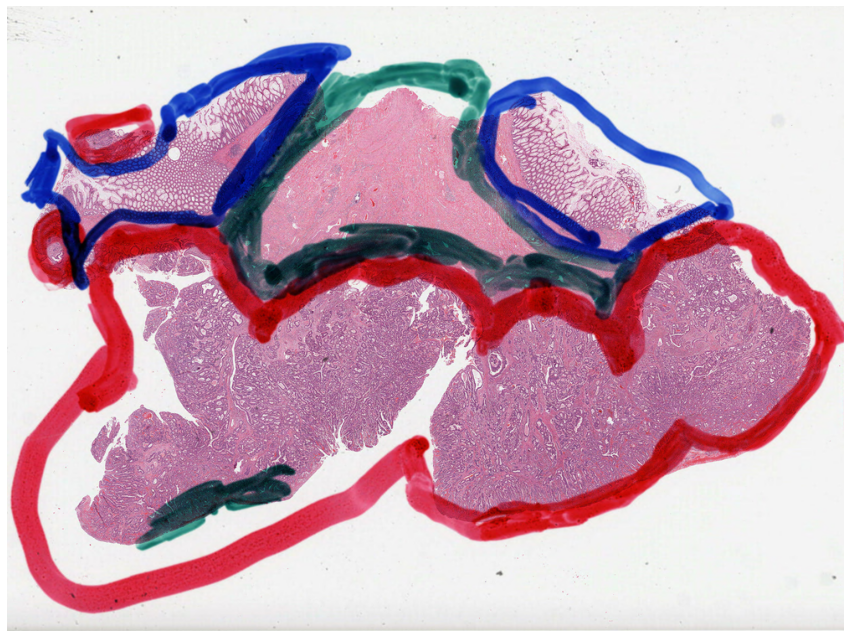}
         \caption{}
         \label{fig:coad_slide}
     \end{subfigure}
     \hfill
     \begin{subfigure}[b]{0.30\textwidth}
         \centering
         \includegraphics[width=\textwidth]{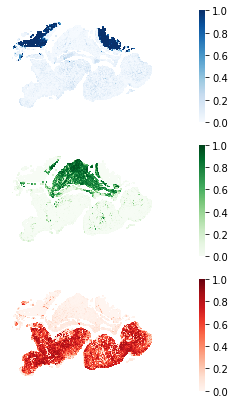}
         \caption{}
         \label{fig:coad_all_clusters}
     \end{subfigure}
}
  \caption{(a) A slide from TCGA-COAD with rough marker annotations. Blue: mucosa with normal intestinal glands, green: muscularis mucosae and submucosa, red: tumoral tissue (b) For each color, the best matching cluster on MoCo v2 features among 10 clusters is displayed.}
\label{fig:coad_cluster}
\end{figure}

In this section, we investigate further why the representation of the feature extractor trained with MoCo v2 is better than its ImageNet counterpart, and propose to apply a clustering algorithm to the tile representations.
We run K-Means (k=10) on the Camelyon16 training tiles features.
We find that one of the 10 clusters is highly correlated with the tumoral tiles.
As a qualitative test, we take the first slide of the Camelyon16 test set and display in Figure \ref{fig:camelyon_cluster} the heatmap of the cluster correlated with the tumoral signal for both ImageNet and MoCo v2 features.
We can see how the best unsupervised MoCo v2 cluster for Camelyon16 tiles almost perfectly matches the tumor annotations, while the ImageNet heatmap only covers a portion of the tumor.

We can evaluate quantitatively this correlation using the local annotations provided within the Camelyon16 dataset.
To do so, we generate a similarity ranking per cluster by computing the cosine distance between their features and each centroid.
We then evaluate this ranking against the tumoral annotations.
The best MoCo v2 cluster achieves 95.1\% AUC on detecting tumoral tiles, while the same method on ImageNet-encoded tiles only achieves 69.4\% AUC. 
Without any supervision, our MoCo v2 features learn to reliably cluster tumoral tiles.

To test if this method generalizes to other datasets, we apply the same clustering principle to the tiles of TCGA-COAD.
Some slides in the dataset contain rough marker annotations, and we display one of them in Figure \ref{fig:coad_slide}.
Two expert pathologists identified the blue region as mucosa with normal intestinal glands, the green region as muscularis mucosae and submucosa, and the red region as tumoral tissue.
Among the 10 tile clusters on MoCo v2 features, we find 3 clusters that match almost perfectly the 3 regions of interest annotated in the slide, as displayed in Figure \ref{fig:coad_all_clusters}.
We were unable to find meaningful clusters on ImageNet encoded tiles.

The five most representative tiles of each cluster can be seen in Supplementary Figure \ref{fig:cam16_top_tiles} for Camelyon16 and Supplementary Figure \ref{fig:coad_top_tiles} for TCGA-COAD.

\subsection{Transfer learning of the feature extractor across datasets}
\label{sec:experiments_transfer_learning}

\begin{table}[htbp]
  \centering
  \caption{Cross-validation performance (AUC percentage) when switching feature extractors on Camelyon16 and TCGA-COAD. For each MIL architecture, we display the performance without transfer learning (on the diagonal) and the performance with transfer of the feature extractor from the other dataset (off diagonal).}
  \label{tab:transfer_table}
  \begin{tabular}{llS[table-format=1.1(2)]S[table-format=1.1(2)]}
  \toprule
  \bfseries \shortstack{MIL\\Architecture} & \bfseries \shortstack{MoCo v2\\Trained On} & \bfseries \shortstack{Metastasis \\ Detection} & \bfseries \shortstack{CMS \\ Classification} \\
  \midrule
  Weldon & Camelyon16 & 97.9(14) & 68.5(48)  \\
         & TCGA-COAD  & \boldentry{1.1(2)}{98.2(10)} & \boldentry{1.1(2)}{80.1(33)}  \\
  \midrule
  Chowder & Camelyon16 & 98.3(12) & 67.4(43)  \\
          & TCGA-COAD  & \boldentry{1.1(2)}{98.6(9)} & \boldentry{1.1(2)}{78.3(36)}\\

  \midrule
  DeepMIL & Camelyon16  & \boldentry{1.1(2)}{96.3(21)} &  74.5(37)  \\
          & TCGA-COAD   & 94.7(39) &  \boldentry{1.1(2)}{80.0(35)}  \\
  \bottomrule
  \end{tabular}
\end{table}

One drawback of our method is that the training of the self-supervised learning algorithm requires a lot of GPU time.
Our Camelyon16 encoder took 13 GPU days to train on 2.6 million tiles.
Retraining a new feature extractor for every dataset is not always feasible. 

In this section, we investigate how well a feature extractor trained using MoCo v2 is able to transfer across datasets.
We benchmark the encoder trained on Camelyon16 on the Consensus Molecular Subtype classification task, and reciprocally the encoder trained on TCGA-COAD on the tumoral classification task.
For each task, we do 5 repeated 5-fold cross-validation, resulting in 25 independent runs.

In Table \ref{tab:transfer_table}, we observe that the feature extractor trained on TCGA-COAD transfers very well to Camelyon16, even outperforming by a small margin the original Camelyon16 feature extractor for the Weldon and Chowder architectures.
However, the feature extractor trained on Camelyon16 introduces a significant drop in performance on the CMS prediction task compared to the TCGA-COAD feature extractor.
On the CMS classification task, the results of the Camelyon16 feature extractor are on par with results from the ImageNet encoder (Table \ref{tab:coad}). 
Campanella et al. \cite{Campanella2019Clinical-gradeImages} experienced a similar drop in performance ($-20.2$ AUC points) when transferring a breast cancer metastasis detection model trained on Camelyon16 to an external real-world dataset.

To explain that difference, we hypothesize that the feature extractor trained on TCGA-COAD was able to capture a much more robust representation, as TCGA-COAD contains slides coming from 24 different centers and is closer to a real-world dataset with variations in slide preparation, staining and scanning.
On the other hand, the Camelyon16 cohort only comes from 2 different centers and is well curated, which reduces variability.
Furthermore, colorectal cancer exhibits a greater variety of histological patterns compared to metastatic lymph nodes.

\section{Conclusion}
\label{sec:conclusion}

Using self-supervision to train the feature extractor, our generic pipeline considerably increases the performance of weakly-supervised learning across three architectures and two datasets, and closes the gap with strongly-supervised models on Camelyon16.
An in-domain feature extractor trained with MoCo v2 can therefore act as a drop-in replacement for any histology model currently relying on ImageNet pre-trained networks.
Furthermore, the learned embedding space clusters histology tiles into biologically meaningful histological patterns, which could lead to new interactive tools for pathologists to explore WSIs and find novel biomarkers.
Finally, by showing transfer learning capabilities, the proposed pipeline lays the foundation for a universal self-supervised histology feature extractor on H\&E-stained slides.

\begin{ack}
We thank Mathieu Andreux, Eric W. Tramel, Charlie Saillard, Alberto Romagnoni, Pierre Manceron, and Aurélie Kamoun for their helpful comments.

We thank the two pathologists Julia Leiner and Vincent Jehanno for validating the results of our clustering model.

This work was granted access to the HPC resources of IDRIS under the allocation 2020-AD011011731 made by GENCI.

The results published here are in part based upon data generated by the TCGA Research Network: \url{https://www.cancer.gov/tcga}.
\end{ack}

\medskip

\small

\clearpage

\bibliography{references}

\clearpage

\appendix

\setcounter{table}{0}
\renewcommand{\tablename}{Supplementary Table}
\renewcommand{\thetable}{\arabic{table}}
\renewcommand{\theHtable}{Supplementary\thetable}

\setcounter{figure}{0}
\renewcommand{\figurename}{Supplementary Figure}
\renewcommand\thefigure{\arabic{figure}}
\renewcommand{\theHfigure}{Supplementary\thefigure}

\section{Implementation details for MIL algorithms}
\label{apd:mil}

\paragraph{Weldon}

A set of one dimensional embeddings is computed for the tile features using a multi-layer perceptron (MLP) with 128 hidden neurons.
We select R=5 top and bottom scores and average them.

\begin{table}[htbp]
  \centering
  \caption{Weldon implementation details}
    \label{tab:weldon_params}

    \begin{tabular}{|l|c|}
    \hline
    \bfseries Layer & \bfseries Type\\\hline
    1 & fc-128\\
    2 & fc-1\\
    3 & extreme-scores-5\\
    4 & average + sigmoid\\\hline
    \end{tabular}
\end{table}

\paragraph{Chowder}

We use the same parameters as in Courtiol et al. \cite{Courtiol2018ClassificationApproach}.
A set of one dimensional embeddings is computed for the tile features.
We select R=5 top and bottom scores and apply a MLP with 200 and 100 hidden neurons and sigmoid activations to the results.

\begin{table}[htbp]
\centering
  \caption{Chowder implementation details}
    \label{tab:chowder_params}

    \begin{tabular}{|l|c|}
    \hline
    \bfseries Layer & \bfseries Type\\\hline
    1 & fc-1\\
    2 & extreme-scores-5\\
    3 & fc-200 + sigmoid\\
    4 & fc-100 + sigmoid\\
    5 & fc-1 + sigm\\\hline
    \end{tabular}
\end{table}

\paragraph{DeepMIL}

A linear layer with 128 neurons is applied to the embedding followed by a Gated Attention layer with 128 hidden neurons. We then apply a MLP with 128 and 64 hidden neurons and ReLU activations to the results.

\begin{table}[htbp]
\centering
  \caption{DeepMIL implementation details}
    \label{tab:deep_mil_params}
  
    \begin{tabular}{|l|c|}
    \hline
    \bfseries Layer & \bfseries Type\\\hline
    1 & fc-128\\
    2 & gated-attention-128\\
    3 & fc-128 + relu\\
    4 & fc-64 + relu\\
    3 & fc-1 + sigmoid\\\hline
    \end{tabular}
  
\end{table}

\paragraph{Tiling}

In our experiments, we extract from each WSI a maximum of 10,000 tiles of size $224 \times 224$ at a zoom level of 0.5 microns per pixel. When the WSI contains less than 10,000 tiles we take all tiles available, otherwise we randomly sample 10,000 tiles.

\paragraph{Optimization details}

We train for 15 epochs with a batch size of 16 using Adam with a learning rate of 0.001, betas=(0.9, 0.999), epsilon=1e-8 and no weight decay.

\newpage

\section{Implementation details for MoCo v2}
\label{apd:moco}

\begin{table}[htbp]
\centering
  \caption{MoCo v2 data augmentation details}
  \label{tab:mocov2_da}

    \begin{tabular}{|l|l|}
    \hline
    \bfseries Type & \bfseries Parameters\\\hline
    RandomRotate & 90\degree, 170\degree, 280\degree\\
    RandomVerticalFlip & -\\
    RandomHorizontalFlip & -\\
    RandomResizedCrop & scale=(0.2, 1.0)\\
    ColorJitter & brightness=0.8\\ 
                & contrast=0.8\\
                & saturation=0.8\\
                & hue=0.2\\
    RandomGrayscale & -\\
    GaussianBlur & sigma\_min=0.1\\
    & sigma\_max=2.0\\\hline
    \end{tabular}
  
\end{table}

\paragraph{Training dataset}
For each WSI in the cohort, we extract a maximum of 10,000 tiles of size $224 \times 224$ at a zoom level of 0.5 microns per pixel. When the WSI contains less than 10,000 tiles we take all tiles available, and we otherwise randomly sample 10,000 tiles.
The training dataset on Camelyon16 contains 2,646,426 million tiles, while the one on TCGA-COAD contains 2,486,159 million tiles.

\paragraph{Data augmentation}
The data augmentation pipeline used during training is described in Supplementary Table \ref{tab:mocov2_da}.

\paragraph{Optimization} We train for 200 epochs with a batch size of 4,096 over 16 NVIDIA Tesla V100.
We use the LARS optimizer with a learning rate of 0.2, momentum of 0.9, weight decay of 1.5e-6 and eta of 1e-3.

\newpage

\begin{figure}[htbp]
\centering

  \includegraphics[width=.77\textwidth]{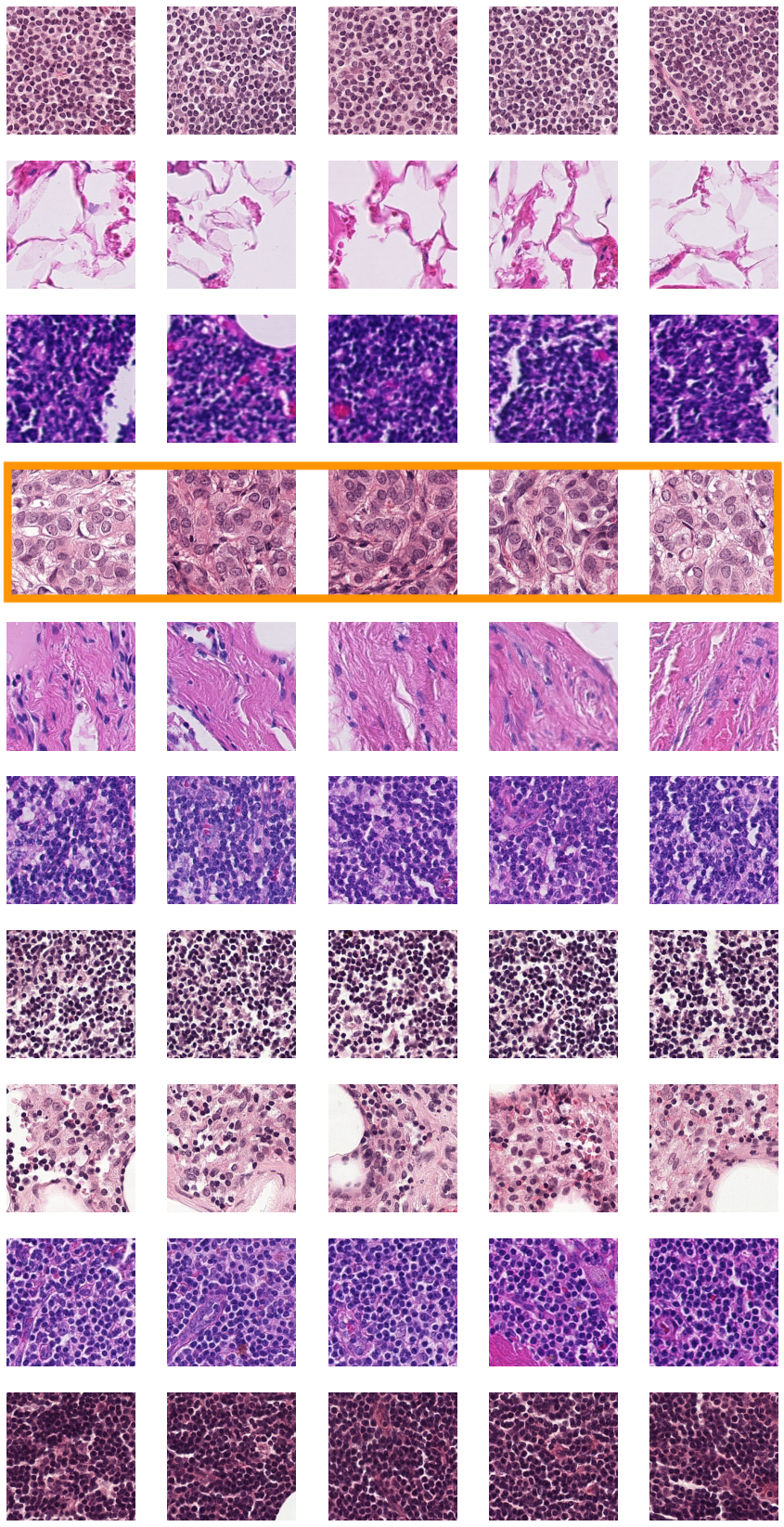}
  \caption{The five most representative tiles of each of the 10 clusters found in the Camelyon16 tile embedding for MoCo v2. In orange, tumoral tissue. Please refer to Figure \ref{fig:cam_test_001_moco_heatmap} for an example of this cluster overlayed on a Camelyon16 slide.}
  \label{fig:cam16_top_tiles}
\end{figure}

\begin{figure}[htbp]
\centering

  {\includegraphics[width=.77\textwidth]{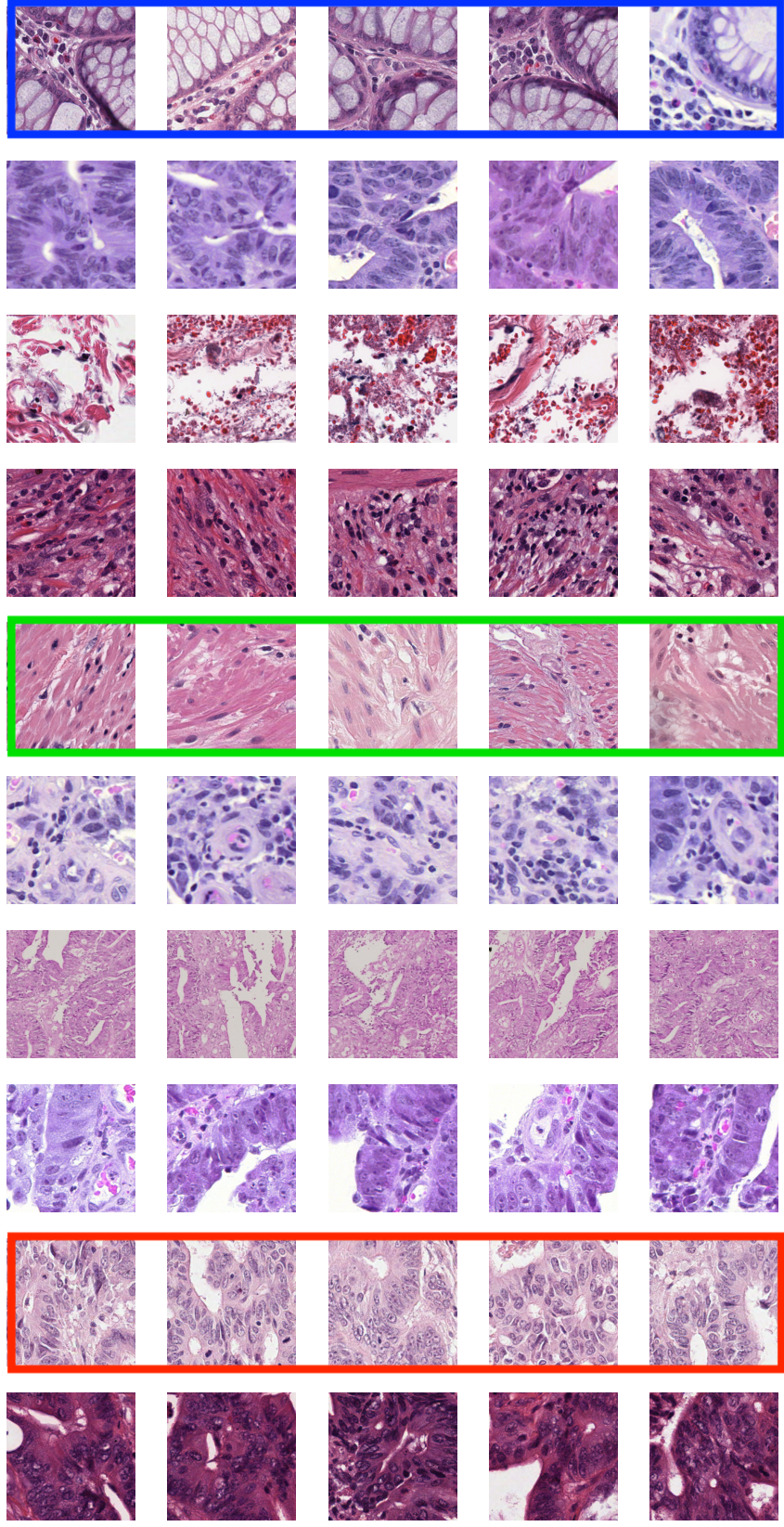}}
  
    \caption{The five most representative tiles of each of the 10 clusters found in the TCGA-COAD tile embedding for MoCo v2. In blue, mucosa with normal intestinal  glands. In green, muscularis mucosae and submucosa. In red, tumoral tissue. Please refer to Figure \ref{fig:coad_all_clusters} for an example of these three clusters overlayed on a TCGA-COAD slide.}
  \label{fig:coad_top_tiles}
\end{figure}

\end{document}